\documentclass[epj]{svjour}
\usepackage{graphicx}
\begin{document}
\title{A theoretical view on bound antikaon-nuclear states}
\author{A. Ramos\inst{1} \and V. K. Magas\inst{1}
\and E. Oset\inst{2} \and H. Toki\inst{3}
}                     
%
%
\institute{Departament d'Estructura i Constituents de la Mat\`eria,
Universitat de Barcelona,
 Diagonal 647, 08028 Barcelona, Spain \and
 Departamento de F\'{\i}sica Te\'orica and IFIC Centro Mixto
Universidad de Valencia-CSIC, Institutos de Investigaci\'on de Paterna \\
Apdo. correos 22085, 46071, Valencia, Spain
\and
Research Center for Nuclear Physics, Osaka University,
Ibaraki, Osaka 567-0047, Japan
}
\date{Received: date / Revised version: date}
%
\abstract{ We present an overview of the latest theoretical
studies on the antikaon properties in the nuclear medium, in
connection with the recent experimental claims of very deeply
bound antikaon-nuclear states. We argue that proper many-body
formulations using modern realistic antikaon-nucleon interactions
are not able to generate such systems. Instead, a simple
two-nucleon antikaon absorption mechanism where the remaining
nucleus acts as spectator explains the peak in the semi-inclusive
proton momentum spectrum, observed on a $^4$He target at KEK (but
later not confirmed in an inclusive experiment) and on a $^6$Li
target at FINUDA. This signal is clearly seen in another FINUDA
experiment measuring the invariant mass of $\Lambda$-proton pairs
after two-nucleon kaon absorption. We show that another peak
of this experiment, seen at lower invariant masses and interpreted
as a bound $K^- pp$ state, is simply generated by the same
two-nucleon absorption mechanism followed by final-state
interactions of the produced particles with the residual nucleus.
Our conclusion is that all the experimental claims for the
formation of very deeply bound antikaonic nuclear systems receive
an alternative explanation in terms of conventional nuclear
processes.
} 
\maketitle
\section{Introduction}
\label{intro}
Over the last years, the theoretical study of the interactions of antikaons
with nuclei has
received a lot of attention due to the important implications on the feasible
realization of interesting physics phenomena.
The possible appearance of a condensate of
antikaons in neutron stars was postulated after examining the
sizable attractive properties of the chiral ${\bar K}N$ interaction at
tree level \cite{kaplan-nelson}. Phenomenological fits to kaonic
atoms fed the idea because a solution were antikaons would feel
strongly attractive potentials of the order of $-200$ MeV
 at the center of the nucleus
\cite{friedman-gal} was preferred. However,
a more satisfactory theoretical understanding of the size of the
antikaon optical potential demands it to be linked to the
elementary ${\bar K}N$ scattering amplitude which is dominated by
the presence of a resonance, the $\Lambda(1405)$, located only 27
MeV below threshold. This makes the problem to be a highly non-perturbative
one. In recent years, the scattering of ${\bar K}$ mesons with nucleons
has been treated within the context of chiral unitary methods
\cite{kai95,oset98,oller01,lutz02,garcia03,jido03,borasoy05,oller05}.
The explicit incorporation of medium effects, such as
Pauli blocking, were shown
to be important \cite{koch} and it was soon realized that
the influence of the resonance demanded the in-medium amplitudes
to be evaluated self-consistently \cite{lutz}.
The resulting antikaon optical potentials
were quite shallower than the
phenomenological one, with depths between $-70$ and $-40$ MeV
\cite{ramos00,schaffner,galself},
but gave an acceptable description of the
kaonic atom data \cite{hirenzaki00,baca00}.

More recently, variational calculations of few body systems using
a simplified phenomenological ${\bar K}N$ interaction predicted
extremely deep kaonic states in $^3$He and $^4$He, reaching
densities of up to ten times that of normal nuclear matter
\cite{akaishi02,dote04,akaishi05}. Motivated by this finding, experiment KEK-E471
 used the
$^4$He(stopped $K^-,p$) reaction and reported \cite{suzuki-kek}
 a structure in the proton momentum
spectrum, which was denoted as the tribaryon $S^0(3115)$ with strangeness
$S=-1$. If interpreted as a $(K^-pnn)$ bound state, it would
have a binding energy of 194 MeV. However, in a recent work
\cite{oset-toki} strong criticisms to the theoretical
claims of Refs.~\cite{akaishi02,dote04,akaishi05} have been put forward and,
in addition, a reinterpretation
of the KEK proton peak has been given in terms of two nucleon absorption
processes, $K^- p n \to \Sigma^- p, K^- p p \to
\Sigma^0(\Lambda) p$, where the rest of the nucleus acts as a
spectator. We note that deficiencies in the
efficiency corrections were
revealed in the last HYP06 conference, and the new reanalyzed KEK
proton momentum spectrum does not show
evidence of this peak \cite{iwasaki}, although the conditions
of the analysis (cuts and coincidences with other charged
particles) were not the same as in the previous one. In contrast, a
FINUDA proton momentum spectrum from $K^-$ absorption on
$^6$Li \cite{finuda_np06} does show a peak, which is interpreted with
the two-nucleon absorption mechanism advocated in
Ref.~\cite{oset-toki}.

Another experiment of the FINUDA collaboration has measured the
invariant mass distribution of $\Lambda p$ pairs \cite{finuda}.
The spectrum shows a narrow peak at 2340 MeV, which corresponds to
the same signal seen in the proton momentum spectrum, namely $K^-$
absorption by a two-nucleon pair leaving the daughter nucleus in
its ground state. Another wider peak is also seen at around 2255
MeV, which is interpreted in Ref.~\cite{finuda} as being a $K^-
pp$ bound state with $B_{K^- pp} = 115^{+6}_{-5}({\rm
stat})^{+2}_{-3}({\rm syst})$ MeV and having a width of $\Gamma =
67^{+14}_{-11}({\rm stat})^{+2}_{-3}({\rm syst})$ MeV. In a recent
work \cite{magas06} we showed that this peak is generated from the
interactions of the $\Lambda$ and nucleon, produced after $K^-$
absorption, with the residual nucleus. We here summarize the
present status of the theoretical studies on the antikaon
properties in the nuclear medium, in connection with the recent
experimental claims of very deeply bound antikaon-nuclear states.

\section{Kaon-nucleon interaction and kaon-nucleus
optical potential}

The existence of deeply bound kaonic states is linked to the
properties of the antikaon-nucleon (${\bar K}N$) interaction and
its modification due to the nuclear medium. In particular, the
understanding of the $\Lambda(1405)$, which is 27 MeV below the
${\bar K}N$ threshold and decays exclusively into $\pi\Sigma$
states, is essential in order to study the fate of the kaons in
the nucleus. The properties of the $\Lambda(1405)$ were nicely
reproduced long ago from the solution of the Schr\"odinger
equation in two (${\bar K}N,\pi\Sigma$) coupled channels
\cite{dalitz}. The model also reproduced a repulsive $K^- p$
amplitude at threshold, a feature which is well known both from
extrapolation of scattering data \cite{martin}, and from the
measurement of kaonic hidrogen at KEK \cite{kek} and DA$\Phi$NE
\cite{dear}.

All the features discussed above are worked out more
systematically in the SU(3)$_f$ chiral unitary models
\cite{kai95,oset98,oller01,lutz02,garcia03,jido03,borasoy05,oller05},
where one uses input from chiral Lagrangians and, in addition to
the $\bar{K} N, \pi \Sigma$ channels, one also has other channels
from the combination of the octet of pseudoscalar mesons with the
octet of stable baryons.  The chiral models generate the
$\Lambda(1405)$ and reproduce various $K^-p$ reactions. In
addition, the $\Lambda(1405)$ is revealed as being the
superposition of two poles of the scattering matrix, one appearing
around 1385 MeV, with a width of 140 MeV, and coupling mostly to
$\pi\Sigma$, the other appearing around 1420 MeV, with a 30 MeV
width and coupling mostly to ${\bar K}N$. This is corroborated by
the analysis of the $K^- p \to \pi^0 \pi^0 \Sigma^0$ data
\cite{prakhov} using the chiral model amplitudes \cite{magas05}.

\begin{figure}
\begin{center}
 \includegraphics[width=7.5cm]{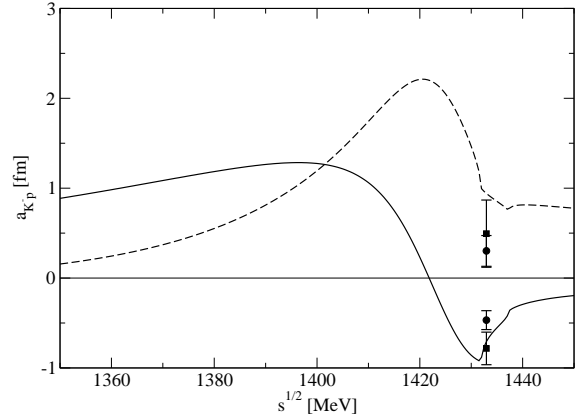}
\end{center}
\caption{Real (solid) and imaginary (dotted) parts
of the $K^- p$ scattering amplitude, as functions of the ${\bar K}N$ center of
mass energy $\sqrt{s}$. The real and imaginary parts of the
scattering length measured at KEK \cite{kek} (squares) and
DA$\Phi$NE \cite{dear} (dots)
are also displayed.
}
\label{fig:scat}       
\end{figure}

The procedure to obtain the meson-baryon amplitude in the chiral
unitary models is extremely simple. From the lowest order chiral
lagrangian for the meson baryon interaction,
\begin{equation}
L_1^{(B)} = \langle \bar{B} i \gamma^{\mu} \frac{1}{4 f^2}
[(\Phi \partial_{\mu} \Phi - \partial_{\mu} \Phi \Phi) B
- B (\Phi \partial_{\mu} \Phi - \partial_{\mu} \Phi \Phi)] \rangle \ ,
\end{equation}
where $\Phi$ and $B$ are the ordinary SU(3) matrices of the meson
and baryon fields respectively,  one deduces the s-wave kernel of
the Bethe-Salpeter equation (potential V):
\begin{equation}
V_{i j} = - C_{i j} \frac{1}{4 f^2}(2\sqrt{s} - M_{i}-M_{j})
\sqrt{\frac{M_{i}+E_i}{2M_{i}}} \sqrt{\frac{M_{j}+E_j}{2M_{j}}}
 \ ,
\label{eq:ampl2}
\end{equation}
where $M_i$ and $E_i$  denote the mass and energy of the baryon in
channel $i$. The coefficients of the symmetric matrix $C_{i j}$
are given in \cite{oset98}. The scattering amplitude is then
obtained by solving the coupled-channel Bethe-Salpeter equation,
\begin{equation}
T = [1 - V \, G]^{-1}\, V \ ,
\end{equation}
using on-shell amplitudes.
The loop function $G$ must be regularized by means of either a
cut-off or by a subtraction method in dimensional regularization.
The real and imaginary parts of the $K^-p$ scattering length,
\begin{equation}
a_{K^- p} = - \frac{1}{4\pi} \frac{M_p}{\sqrt{s}} T_{K^- p \to K^- p} \ ,
\label{eq:scatlength}
\end{equation}
obtained from the lowest order chiral model of Ref. \cite{oset98},
but using the dimensional regularization scheme described in
\cite{jido03}, are displayed in Fig.~\ref{fig:scat}. Note that the
real part is negative (repulsive) at threshold and agrees with the
KEK data \cite{kek} but not with the recent DEAR determination
\cite{dear}, which is about 30\% smaller. A recent model,
including also the next to leading order terms in the chiral
kernel \cite{borasoy05} achieves a better agreement with the real
part of the DEAR scattering length, but still misses the imaginary
part by about half the size.

One needs next to implement the corrections to the $\bar{K} N$
amplitude in the medium. Pauli blocking acting on the intermediate
loop nucleons forces them to be placed above de Fermi momentum,
which costs more energy and induces a repulsive effect on the
resonance. Correspondingly, the $K^- p$ amplitude at threshold
changes from being repulsive to being attractive, providing in
this way a source of attraction for the ${\bar K}$ self-energy or
optical potential. One should not stop here since the antikaons in
the intermediate states feel this attractive potential. It then
costs less energy to excite them, which partly compensates the
repulsive Pauli blocking effect. As a consequence, the resonance
moves back to lower energies and the new kaon self-energy will not
be as attractive. This self-consistency procedure was done for the
first time in Ref. \cite{lutz}. It was found that the attraction
felt by the kaons was drastically reduced. Subsequent calculations
\cite{schaffner,galself}, including those which consider also the
renormalization of the intermediate pions in the $\pi\Sigma$
channel \cite{ramos00} were performed and the strength of the
potential was shown to be attractive and of the order of 40-70 MeV
at nuclear matter density.

\begin{figure}
\begin{center}
 \includegraphics[width=5cm]{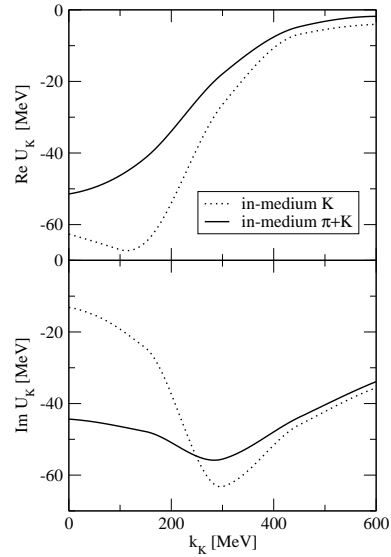}
\end{center}
\caption{Real (top) and imaginary (bottom) part of the $K^-$
optical potential as a function of momentum obtained from two
approximations: {\it in-medium K} (dotted lines): only antikaons
are (self-consistently) dressed and {\it in-medium $\pi+$K} (solid
lines): pions and antikaons are dressed. } \label{fig:opt}
\end{figure}

Indeed, from the self-consistent ${\bar K}N$ amplitudes obtained
following the model of Ref. \cite{ramos00}, improved by including
the p-wave components of the kaon self-energy renormalized by the
effect of short-range correlations \cite{tolos06}, one obtains the
antikaon optical potential displayed in Fig.~\ref{fig:opt} as a
function of the antikaon momentum at normal nuclear density,
$\rho=0.16$ fm$^{-3}$. The real part of the optical potential,
evaluated at the corresponding quasi-particle energy,
$\varepsilon_{K} = k_{K}^2/(2 m_K) + {\rm Re}\,U_K$, is attractive
and around $-50$ MeV at zero momentum. The attraction is somewhat
larger if the dressing of pions in the intermediate loops is
ignored, as done in most calculations. It is important to notice
that the self-consistent procedure, including also the p-wave
excitation of $\Lambda$-hole and $\Sigma$-hole components in the
antikaon self-energy, automatically incorporates the absorption of
the antikaons by two nucleons, namely $K^- NN \to N \Lambda, N
\Sigma$, in the imaginary part of the optical potential. The
dressing of the pion brings an additional source of absorption
strength through this processes but also through new channels (via
the $\Delta$-hole pionic excitations),
 $K^- NN \to \pi N \Lambda, \pi N \Sigma$, which turn out to be very relevant
 in the low momentum region of the kaon optical potential.

This potential, evaluated at $k_K=0$ MeV and $\varepsilon_K=m_K$,
provides a good reproduction of the kaonic atom data in light and
medium heavy nuclei \cite{hirenzaki00}, with small diversions that
were quantified in a best fit analysis \cite{baca00}. We can then
conclude that the SU(3) chiral unitary model with the inclusion of
all the medium corrections on the baryons and the mesons is
supported from the kaonic atom data.

With this behavior for the ${\bar K}N$ dynamics in nuclei having
been established, the claim given in \cite{akaishi02,dote04} on
the possible existence of narrow deeply bound states in light
nuclei, predicting $A=3$ $I=0$ and $I=1$ states with binding
energies of around $100$ MeV, was surprising. The situation became
more puzzling when an experiment searching for deeply bound $K^-$
states using the reaction $^4$He(stopped $K^- , p)$
\cite{suzuki-kek} found a signal for a $I=1$ strange tri-baryon,
$S^0(3115)$, since its interpretation as a bound state would imply
a binding energy around $200$ MeV. Subsequent corrections
including relativistic effects \cite{akaishi05}, supplied the
necessary amount of attraction, yielding a ${\bar K}$-nucleus
potential strength of $-600$ MeV and nuclear densities at the center
of the nucleus of about 9 times nuclear matter saturation density.
A closer look at the model interaction and the many-body
approximations employed in these works reveals serious
deficiencies that were pointed out in Ref.~\cite{oset-toki}. See
also the criticisms put forward in Ref.~\cite{weisehere}. Among
those: 1) The theoretical potential in
\cite{akaishi02,dote04,akaishi05} eliminates the direct coupling
of $\pi \Sigma \rightarrow \pi \Sigma$ and $\pi \Lambda
\rightarrow \pi \Lambda$ in contradiction with the results of
chiral theory that establishes large couplings for them
\cite{oset98}. 2) Self-consistency, shown to be essential in
regulating the size of attraction felt by the antikaons, was not
implemented. 3) Finally, short range correlations, which would
prevent reaching the high densities found, were not incorporated
realistically.

An important turnover in the understanding of deeply bound kaonic
systems has been given by recent few body calculations applied to
the $K^- p p$ system. Three-body coupled channel Faddeev equations
are solved in Ref.~\cite{shevchenko}, using separable two-body
interactions fitted to scattering data. A $I=1/2$ three-body
${\bar K}(NN)_{I=1}$ state was found with a binding energy in the
range $B \sim 50-70$ MeV and a width $\Gamma \sim 95 -110$ MeV.
The recent variational calculations of Ref.~\cite{dote_hyp06}
implement, as a major change with respect the previous works
\cite{akaishi02,dote04,akaishi05}, realistic short-range NN
correlations, which prevent from reaching the high central
densities obtained before. Although still preliminary, the first
results point at a binding energy for the $K^- pp$ system of
around 50 MeV and a total width for this state larger than 100
MeV, the absorption $K^- N N \to Y N$ channel, not considered in
\cite{shevchenko}, contributing by about 20\%.

\section{Proton momentum spectrum}

Convinced by the outcome of realistic many-body calculations, both
in nuclear matter and in finite nuclei, that the peak in the
proton momentum spectrum found at KEK \cite{suzuki-kek} could not
correspond to a deeply bound $K^-$ state, the authors of
Ref.~\cite{oset-toki} searched for other explanations and found a
natural mechanism that passed all tests. The peak is due to the
two-body absorption mechanism, $K^- NN \rightarrow \Sigma p$, in
$^4$He leaving the other nucleons as spectators. The candidate
reaction to produce a peak is $K^- {}^4{\rm He} \to \Sigma^- p d$,
which has a production rate of 1.6\% \cite{katz}. This is
compatible with demanding that 1\% of the strength proceeds
through the $d$ acting as a spectator, as required by the two-body
absorption mechanism, which is the formation branching ratio per
stopped $K^-$ for the peak seen in Ref.~\cite{suzuki-kek} just
below 500 MeV/c. The two nucleon absorption mechanism produces a
well defined narrow peak at $p_p \sim 480$ MeV/c, as in the
experiment, if the Fermi motion of the spectator $d$ is ignored,
which is equivalent in this mechanism to ignoring the
corresponding recoil momentum.  These effects have been 
shown \cite{oset-hyp06} to produce a broadening of the proton peak
of $\sim 70$ MeV for $K^-$ absorption in $^4$He and of $\sim 20$
MeV if the absorption takes place in $^6$Li. Actually, the peaks
can be narrower and/or enhanced if some specific kinematical cuts
are applied. For instance, a secondary charged particle is also
detected, together with the proton, in the KEK experiment and the
peak becomes more prominent when this secondary particle is a
$\pi^-$.
We note that the KEK group reported recently \cite{iwasaki} some
deficiencies in correcting the proton efficiencies of their
previous experiment, and the new proton spectrum from the reaction
$^4{\rm He}({\rm stopped}\,K^-,p)$ no longer shows a peak. It
remains to be seen if the peak would reappear if the same
semi-inclusiveness conditions as in the earlier experiment were
applied. What is clear is that recoil-broadening effects of 70 MeV
plus formation rates of $\sim 1$\% make the observation of this
peak hard, unless specific kinematical and particle selection cuts
are applied. This is precisely what was done in the proton
momentum spectrum reported by the FINUDA collaboration from $K^-$
capture on $^6$Li \cite{finuda_np06}, where the peak observed at
around 500 MeV/c was correlated with a $\pi^-$ coming from
$\Sigma^-$ decay in flight, selected by choosing pion momenta
larger than 275 MeV/c. It was also shown there that most of the
events fulfilled the back-to-back condition $\cos\theta_{\pi^- p}
< -0.8$, confirming the origin of the proton peak as coming from
the two-nucleon $K^-$ absorption mechanism. We note that this
interpretation demanded the peak corresponding to the reaction
$K^- NN \rightarrow \Lambda p$ be seen as well, which was indeed
the case in the FINUDA experiment \cite{finuda_np06} where a
feeble signal can be observed at about 580 MeV/c.

The signals in the proton spectrum should appear when a proton as
well as a $\Sigma$ or a $\Lambda$ would be emitted back to back
after two-nucleon $K^-$ absorption and a residual nuclear system
remains as a spectator and stays nearly in its ground state. This
last requirement becomes more difficult in heavier nuclei since
the distortion of the $\Lambda$ or $p$ particle in their way out
through the nucleus leads unavoidably to nuclear excitations. As a
consequence of this, one expects the signals to fade gradually for
heavier nuclei, which would explain the smoothness of the proton
spectrum from absorption on $^{12}$C, also measured in
\cite{finuda_np06}.

\section{The $\Lambda p$ invariant mass spectrum}
\label{sec:2}

\begin{figure*}
\begin{center}
 \includegraphics[width=7cm,height=14cm,angle=-90]{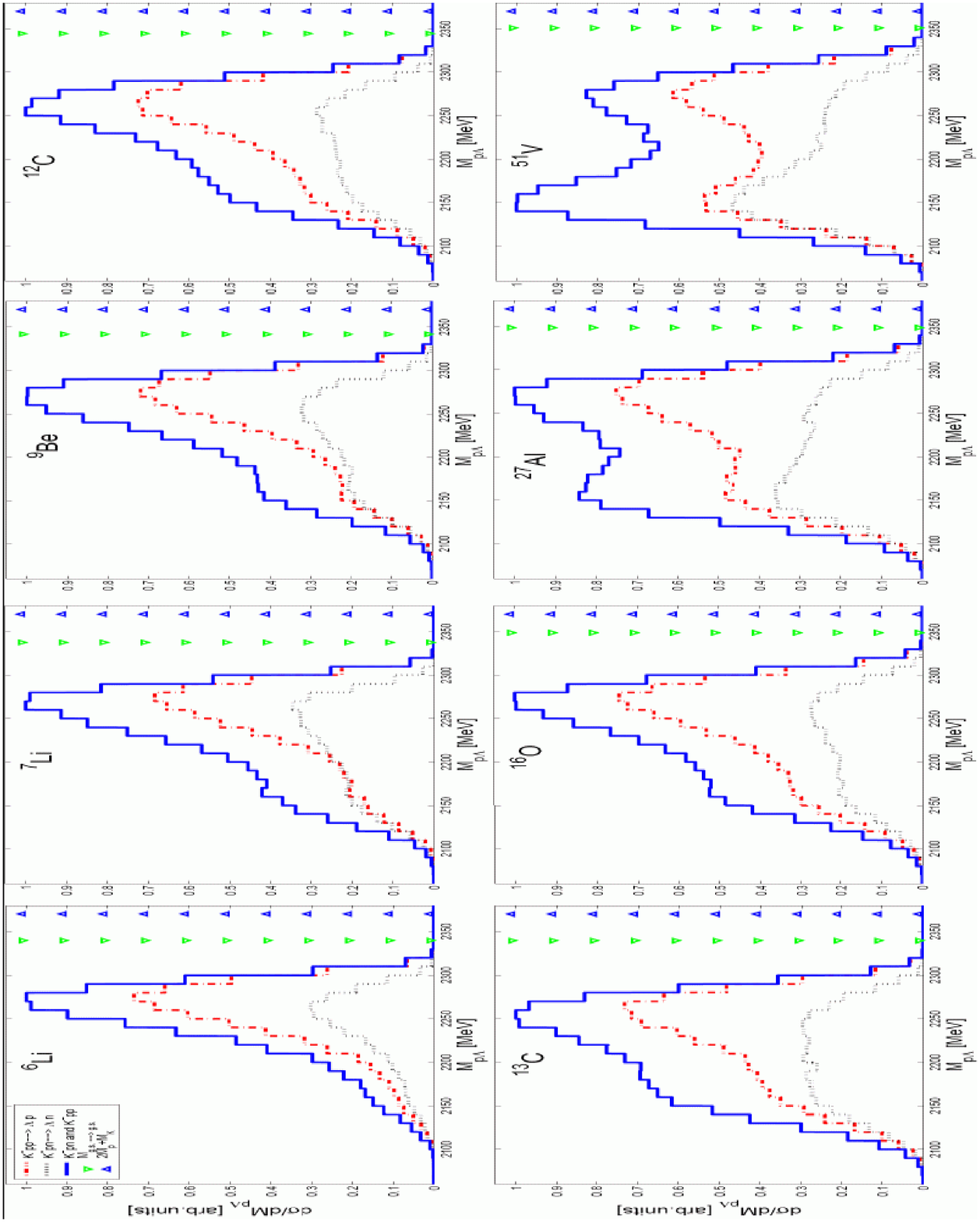}
\end{center}
\caption{$\Lambda p$ invariant mass distribution after $K^-$
absorption in several nuclei imposing
 $P_\Lambda > 300$ MeV/c and
$\cos \Theta_{\vec{p}_\Lambda \vec{p}_p}<-0.8$. }
\label{fig:1}       
\end{figure*}

Having thus reinterpreted the alleged deeply bound kaonic state,
reported from the proton spectrum experiment E471 at KEK, as
simply coming from a two-nucleon $K^-$ absorption mechanism, a
reanalysis of another FINUDA experiment, claiming to have formed a
$K^- p p$  state bound by $115$ MeV seen in a $\Lambda p$
invariant mass spectrum \cite{finuda}, is an absolute necessity.
This was done in Ref.~\cite{magas06} and, in this section, we
summarize the main ideas and results of that work, implementing
some minor improvements.

The FINUDA collaboration in \cite{finuda} looked for $\Lambda p$
events back to back following the $K^- p p$ absorption in $^6$Li,
$^7$Li and $^{12}$C nuclei. We have developed a simulation code
for the reaction $ K^- ~ A \rightarrow \Lambda ~ p ~ A'$
  with stopped kaons, which accounts for the final state interaction of the
  produced $\Lambda$ and $p$ with the residual nuclear state. Once this is done,
  the $\Lambda p$
events are collected considering the cuts applied in the experiment:
one selecting $p_\Lambda > 300$ MeV/c, to eliminate events
from $K^- p \rightarrow \Lambda \pi$, and another one imposing
$\cos \Theta_{\vec{p}_\Lambda \vec{p}_p}<-0.8$, to filter $\Lambda
p$ pairs going back to back.

More specifically,
the reaction
$(K^-)_{\rm stopped}  A \rightarrow
\Lambda  p  A'$ proceeds by capturing a slow $K^-$ in
a high atomic orbit of the nucleus, which later cascades down till
the $K^-$ reaches a low lying orbit, from where it is finally absorbed.
We assume
that the absorption of the $K^-$ takes
place from the lowest level where the energy shift for atoms has been
measured, or, if it is not measured, from the level where the calculated
shift \cite{hirenzaki00} falls within the measurable range.
%

The width for $K^-$  absorption from $p N$ pairs in a nucleus with mass number $A$
is given, in Local Density Approximation (LDA) by
\begin{equation}
\Gamma_A  \propto  \int d^3 \vec{r} |\Psi_{K^-}(\vec{r})|^2
\int \frac{d^3 \vec{p}_1}{(2\pi)^3}
\int\frac{d^3 \vec{p}_2}{(2\pi)^3}
 \Gamma_m(\vec{p}_1,\vec{p}_2,\vec{p}_K,\vec{r}),
\label{eq2a}
\end{equation}
where $|\Psi_{K^-}(\vec{r})|^2$ is the probability of finding
the $K^-$ in the
nucleus, $|\vec{p}_1|, |\vec{p}_2| <k_F(r)$ with
$k_F(r)=\left( 3\pi^2 \rho(r)/2 \right)^{1/3}$ being
the local Fermi momentum and
$\Gamma_m(\vec{p}_1,\vec{p}_2,\vec{p}_K,\vec{r})$ is the in-medium
decay width for the $K^- p N \rightarrow \Lambda N$ process. The structure of
the integrals determining $\Gamma_m$,
\begin{equation}
\Gamma_m \propto \int d^3
\vec{p}_\Lambda d^3 \vec{p}_N \dots
K(\vec p_\Lambda,\vec r) K(\vec p_N,\vec r) \ ,
\end{equation}
allows us to follow
the propagation of the produced nucleon and $\Lambda$ through the
nucleus after $K^-$ absorption via
the kernel $K(\vec p, \vec r)$. The former two equations describe
the process in which a kaon at rest is absorbed by two nucleons
($pp$ or $pn$) within the local Fermi sea emitting a nucleon and
a $\Lambda$. The primary nucleon ($\Lambda$) is
 allowed to re-scatter with
nucleons in the nucleus according to a probability per unit length
given by $\sigma_{N(\Lambda)} \rho(r)$, where $\sigma_{N(\Lambda)}$ is the
experimental $NN(\Lambda N)$ cross section at the corresponding energy,
while in
\cite{magas06} a simpler parameterization for the $\Lambda$, of the type
$\sigma_\Lambda=2\sigma_N/3$ was employed.
The angular distribution of the
re-scattered particles are also generated according to experimental
differential cross-sections.
We note that particles move under the
influence of a mean field potential, of Thomas-Fermi type.
After several possible collisions, one or more
nucleons and a $\Lambda$ emerge from the nucleus and the invariant
mass of all possible $\Lambda p$ pairs, as well as their relative
angle, are evaluated for each Monte Carlo event.
See Ref.~\cite{magas06} for more details.

Absorption of a $K^-$ from a nucleus leaving the final daughter
nucleus in its ground state gives rise to a narrow peak in the
$\Lambda p$ invariant mass distribution, as it is observed in the
spectrum of \cite{finuda}. We note that our local density formalism, in which
the hole levels in the Fermi sea form a continuum of states,
cannot handle properly transitions to discrete states of the
daughter nucleus, in particular to the ground state. For this
reason, we will remove in our calculations those events in which
the $p$ and $\Lambda$ produced after $K^-$ absorption leave the
nucleus without having suffered a secondary collision. However,
due to the small overlap between the two-hole initial state after
$K^-$ absorption and the residual $(A-2)$ ground state of the
daughter nucleus, as well as to the limited
 survival probability for
both the $p$ and the $\Lambda$ crossing the nucleus without any collision,
this strength represents only a
moderate fraction, estimated to be smaller than 15\% in $^{7}$Li \cite{magas06}.
This means in practice that the
 excitation of the nucleus will require the secondary collision of the $p$
 or $\Lambda$ after the $K^-pp$ absorption process, similarly as to what
 happens in $(p,p')$ collisions, where the
strength of the cross section to elastic or bound excited states is very
small compared to that of nuclear breakup producing the quasi-elastic peak.

Our invariant mass spectra requiring at least a secondary
collision of the $p(n)$ or $\Lambda$ after the $K^- pp(np)$
absorption process are shown in Fig.~\ref{fig:1} for several
nuclei,
where we have
applied the same cuts as in the experiment, namely  $P_\Lambda >
300$ MeV/c (to eliminate events from $K^- p \rightarrow \Lambda
\pi$) and $\cos \Theta_{\vec{p}_\Lambda \vec{p}_p}<-0.8$ (to
filter $\Lambda p$ pairs going back-to-back). Actually, the calculated
angular distribution shown in Ref.~\cite{magas06} demonstrates
that, even after collisions, a sizable fraction of the events
appear at the back-to-back kinematics. These events  generate
the main bump at 2260-2270 MeV
in all the $\Lambda p$ invariant mass spectra shown in Fig.~\ref{fig:1},
about the same position as the main
peak shown in the inset of Fig. 3 of \cite{finuda}.
We note that, since one
measures the $\Lambda p$ invariant mass, the main contribution
comes from $K^- p p \rightarrow \Lambda p$ absorption (dot-dashed
lines), although the contribution from the $K^- p n \rightarrow
\Lambda n$ reaction followed by $np\rightarrow pn$ (dotted line)
is non-negligible.
 It is interesting to observe that the width
of the distribution gets broader with the size of the nucleus,
while the peak remains in the same location, consistently to what
one expects for the behavior of a quasi-elastic peak. Let us point
out in this context that the work of \cite{Mares:2006vk} shows
that the possible interpretation of the FINUDA peak as a bound
state of the $K^-$ with the nucleus, not as a $K^- pp$ bound
state, would unavoidably lead to peaks at different energies for
different nuclei. We finally observe that the spectra of heavy
nuclei develop a secondary peak at lower invariant masses due to
the larger amount of re-scattering processes. It is slightly more
pronounced than that shown in our earlier work \cite{magas06}, due
to the use here of a more accurate $\Lambda N$ scattering cross
section.



Finally, we compare our results with those presented in the inset
of Fig. 3 in Ref.~\cite{finuda}, using the three lighter targets
in the same proportion as in the experiment (51\% $^{12}$C, 35\%
$^{6}$Li and 14\% $^{7}$Li). We note that the averaged histogram
is dominated by the $^{12}$C component of the mixture, due mostly
to the larger overlap with the kaon wave function. As we see, our
calculations are in excellent agreement with the measured spectrum
in this range.

\begin{figure}
\hspace*{2cm}\includegraphics[width=4cm,height=5cm]{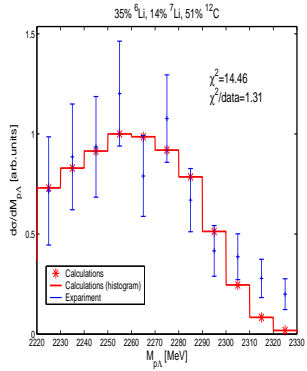}
\caption{(Color online) Invariant mass of $\Lambda p$ distribution
for $K^-$ absorption in light nuclei in the following proportion:
51\% $^{12}$C,  35\% $^{6}$Li and 14\% $^{7}$Li. Stars and
histogram show our results, while experimental points and error
bars are taken from } \label{chi2}
\end{figure}

\section{Concluding remarks}

In summary, we have seen how the experimental $\Lambda p$
invariant mass spectrum of the FINUDA collaboration \cite{finuda}
is naturally explained in our Monte Carlo simulation as a
consequence of final state interactions of the particles produced
in nuclear $K^-$ absorption as they leave the nucleus, without the
need of resorting to exotic mechanisms like the formation of a
$K^- pp$ bound state. Together with the interpretation of the peak
seen in the proton momentum spectrum \cite{suzuki-kek,finuda_np06}
as coming from a two-nucleon kaon absorption mechanism, where the
rest of the nucleus acts as an spectator and the daughter nucleus
remains in its ground state given in
Refs.~\cite{oset-toki,finuda_np06}, it seems then clear that there
is at present no experimental evidence for the existence of deeply
bound kaonic states.

From the theoretical side, the absence of deeply bound kaonic
states is supported by sophisticated many-body calculations of the
optical potential in dense matter
\cite{ramos00,schaffner,galself}. A major step forward has been
achieved by recent few body calculations, either solving Faddeev
equations \cite{shevchenko} or applying variational techniques
\cite{dote_hyp06}, using realistic $\bar{K}N$ interactions and
short-range nuclear correlations. These works predict few-nucleon
kaonic states bound by 50-70 MeV but having large widths of the
order of 100 MeV, thereby disclaiming the findings of
Refs.~\cite{akaishi02,dote04,akaishi05}.


\section{Acknowledgments}
This work is partly supported by contracts BFM2003-00856 and
FIS2005-03142 from MEC (Spain) and FEDER, the Generalitat de
Catalunya contract 2005SGR-00343, and the E.U. FLAVIANET network
contract MRTN-CT-2006-035482. This research is part of the EU
Integrated Infrastructure Initiative Hadron Physics Project under
contract number RII3-CT-2004-506078.


\end{document}